\documentclass[useAMS,usenatbib,  usegraphicx, twocolumn]{mn2e}

\title[Cosmological perturbations in Bose-Einstein condensates]{Evolution of cosmological perturbations in Bose-Einstein condensate dark
matter}

\author[T.  Harko]{T. Harko\thanks{E-mail:
harko@hkucc.hku.hk}
\\
Department of Physics and Center for Theoretical and Computational Physics, The University of Hong Kong, \\
Pok Fu Lam Road, Hong Kong, Hong Kong SAR, P. R. China
}
\begin{document}


\pagerange{\pageref{firstpage}--\pageref{lastpage}} \pubyear{2002}

\maketitle

\label{firstpage}

\begin{abstract}
We consider the global cosmological evolution and the evolution of the density contrast in the Bose-Einstein condensate dark matter model, in the framework of a Post-Newtonian cosmological approach. In the Bose-Einstein  model, dark matter can be
described as a non-relativistic, Newtonian
gravitational condensate, whose density and pressure are
related by a barotropic equation of state. For a
condensate with quartic non-linearity, the equation of state is
polytropic with index $n=1$.The basic equation describing the evolution of the perturbations of the Bose-Einstein condensate is obtained, and its solution is studied by using both analytical and numerical methods. The global cosmological evolution as well as the evolution of the perturbations of the condensate dark matter shows significant differences with respect to the pressureless dark matter model, considered in the framework of standard cosmology. Therefore the presence of condensate dark matter could have modified drastically the cosmological evolution of the early universe, as well as the large scale structure formation process.
\end{abstract}

\begin{keywords}
cosmology: theory: dark matter: large-scale structure of Universe -- instabilities--equation of state
\end{keywords}

\section{Introduction}

Cosmological observations provide compelling evidence that about 95\% of the
content of the Universe resides in two unknown forms of energy that we call
dark matter and dark energy: the first residing in bound objects as
non-luminous matter \citep{dm1, dm2, dm3}, the latter in the form of a zero-point energy
that pervades the whole Universe \citep{PeRa031, PeRa032}. The dark matter is thought
to be composed of cold neutral weakly interacting massive particles, beyond
those existing in the Standard Model of Particle Physics, and not yet
detected in accelerators or in dedicated direct and indirect searches. There
are many possible candidates for dark matter, the most popular ones being
the axions and the weakly interacting massive particles (WIMP) (for a review
of the particle physics aspects of dark matter see \cite{OvWe04}). Their
interaction cross section with normal baryonic matter, while extremely
small, are expected to be non-zero and we may expect to detect them
directly. It has also been suggested that the dark matter in the Universe
might be composed of superheavy particles, with mass $\geq 10^{10}$ GeV \citep{Ko1,Ko2,Ko3, Ko4}. But
observational results show the dark matter can be composed of superheavy
particles only if these interact weakly with normal matter or if their mass
is above $10^{15}$ GeV \citep{AlBa03}. Scalar fields or other long range
coherent fields coupled to gravity have also intensively been used to model
galactic dark matter \citep{Lee96, scal1, scal2, scal3, Ar1, scal4, scal5, scal6, Kh, scal7, Ar2, Ar3, Bri}. Alternative theoretical models to explain the galactic rotation curves have also been elaborated recently \citep{alt1, alt2, alt3, alt4, alt5, alt6, alt7, alt91, alt8,  alt9,  alt10}.

In order to explain the recent observational data, the $\Lambda $CDM ($\Lambda $ Cold Dark Matter) model was developed \citep{PeRa031, PeRa032}. The $\Lambda $CDM  model
successfully describes the accelerated expansion of the Universe, the
observed temperature fluctuations in the cosmic microwave background
radiation, the large scale matter distribution, and the main aspects of the
formation and the evolution of virialized cosmological objects.

Despite these important achievements, at galactic scales $\sim 10$ kpc, the $%
\Lambda $CDM model meets with severe difficulties in explaining the observed
distribution of the invisible matter around the luminous one. In fact, $N$%
-body simulations, performed in this scenario, predict that bound halos
surrounding galaxies must have very characteristic density profiles that
feature a well pronounced central cusp, $\rho _{NFW}(r)=\rho
_{s}/(r/r_{s})(1+r/r_{s})^{2}$ \citep{nfw}, where $r_{s}$ is a scale radius
and $\rho _{s}$ is a characteristic density. On the observational side,
high-resolution rotation curves show, instead, that the actual distribution
of dark matter is much shallower than the above, and it presents a nearly
constant density core: $\rho _{B}(r)=\rho
_{0}r_{0}^{3}/(r+r_{0})(r^{2}+r_{0}^{2})$ \citep{bur}, where $r_{0}$ is the
core radius and $\rho _{0}$ is the central density.

At very low temperatures, all particles in a dilute Bose gas condense to the
same quantum ground state, forming a Bose-Einstein Condensate (BEC), i.e., a
sharp peak over a broader distribution in both coordinates and momentum
space.  A coherent state develops when the particle density is enough
high, or the temperature is sufficiently low. The Bose-Einstein condensation process was first observed experimentally in
1995 in dilute alkali gases, such as vapors of rubidium and sodium, confined
in a magnetic trap, and cooled to very low temperatures. A sharp peak in the
velocity distribution was observed below a critical temperature, indicating
that condensation has occurred, with the alkali atoms condensed in the same
ground state and showing a narrow peak in the momentum space and in the
coordinate space \citep{exp1, exp2, exp3}.
Quantum degenerate gases have
been created by a combination of laser and evaporative cooling techniques,
opening several new lines of research, at the border of atomic, statistical
and condensed matter physics \citep{Da99,rev1,rev2, rev3, rev4, rev5, rev6}.

The possibility that dark matter could be in the form of a Bose-Einstein condensate was considered
 in \citet{Sin1} and \citet{Sin2}. The condensate was described by the non-relativistic
Gross-Pitaevskii equation, and its solution was obtained numerically. An alternative
approach was developed in \citet{BoHa07}. By introducing the Madelung
representation of the wave function, the dynamics of the system can be
formulated in terms of the continuity equation and of the hydrodynamic Euler
equations. Hence dark matter can be described as a non-relativistic,
Newtonian Bose-Einstein gravitational condensate gas, whose density and
pressure are related by a barotropic equation of state. In the case of a
condensate with quartic non-linearity, the equation of state is polytropic
with index $n=1$. To test the validity of the model the Newtonian tangential
velocity equation of the model was fitted with a sample of rotation curves
of low surface brightness and dwarf galaxies, respectively. A very good
agreement was found between the theoretical rotation curves and the
observational data for the low surface brightness galaxies. Therefore dark
matter halos can be described as an assembly of light individual bosons that
acquire a repulsive interaction by occupying the same ground energy state.
That prevents gravity from forming the central density cusps. The condensate
particle is light enough to naturally form condensates of very small masses
that later may coalesce, forming the structures of the Universe in a similar
way than the hierarchical clustering of the bottom-up CDM picture. Then, at
large scales, BEC perfectly mimic an ensemble of cold particles, while at
small scales quantum mechanics drives the mass distribution. Different properties of the Bose-Einstein condensate dark matter halos, like the effects of the rotation and of the vortices, as well as the cosmological effects of the  condensation were also investigated \citep{Fer, Fuk08, Fuk09, Rind, Sik, Lee09, Brook, kai, Lee10}.

It is the purpose of the present paper to study the global cosmological dynamics of
gravitationally self-bound Bose-Einstein dark matter condensates, and the evolution of the small cosmological perturbations in the condensate. The equations of motion of the condensate dark matter are obtained in a Post-Newtonian approximation by using the conservation of the general relativistic energy-momentum tensor, and considering the small velocity limit. The cosmological dynamics of the Bose-Einstein condensate is also studied. The exact solution of the Friedmann equations is obtained, and it is compared with the standard Einstein-de Sitter cosmological model. In order to study the evolution of the small cosmological perturbations the equation describing the Newtonian perturbations with pressure is obtained in a general form, by also taking into account a term which was neglected in the previous studies of this problem  \citep{McCrea51, Har64, Lima97, Reis03, Ab07, Pace10}.  The equation of the density contrast for the Bose-Einstein condensate is investigated by using both analytical and numerical methods. The presence of the condensate dark matter significantly modifies the cosmological dynamics of the Universe, as well as the large scale structure formation.

The present paper is organized as follows. The basic properties of the Bose-Einstein condensate dark matter halos are reviewed in Section \ref{sect2}. The Post-Newtonian hydrodynamical equations of motion for a perfect fluid with pressure are derived in Section \ref{sect3}. The cosmological dynamics of the Bose-Einstein condensate dark matter is considered in Section \ref{sect4}. The equation describing the small cosmological perturbations of a fluid with pressure are derived in Section \ref{sect5}. The evolution of the small cosmological perturbations in a Bose-Einstein condensate is considered in Section \ref{sect6}. We discuss and conclude our results in Section \ref{sect7}.

\section{Dark matter as a Bose-Einstein condensate}\label{sect2}

In a quantum system of $N$ interacting condensed bosons most of the bosons
lie in the same single-particle quantum state. For a system consisting of a
large number of particles, the calculation of the ground state of the system
with the direct use of the Hamiltonian is impracticable, due to the high
computational cost. However, the use of some approximate methods can lead to
a significant simplification of the formalism. One such approach is the mean
field description of the condensate, which is based on the idea of
separating out the condensate contribution to the bosonic field operator. We
also assume that in a medium composed of scalar particles with non-zero
mass, when the medium makes a transition to a
Bose-Einstein condensed phase, the range of Van der Waals-type scalar mediated interactions among
particles becomes infinite.

\subsection{The Gross-Pitaevskii equation}

The many-body Hamiltonian describing the interacting bosons confined by an
external potential $V_{ext}$ is given, in the second quantization, by
\begin{eqnarray}\label{ham}
&&\hat{H}=\int d\vec{r}\hat{\Phi}^{+}\left( \vec{r}\right) \left[ -\frac{\hbar
^{2}}{2m}\nabla ^{2}+V_{rot}\left( \vec{r}\right) +V_{ext}\left( \vec{r}%
\right) \right] \hat{\Phi}\left( \vec{r}\right) +\nonumber\\
&&\frac{1}{2}\int d\vec{r}d%
\vec{r}^{\prime }\hat{\Phi}^{+}\left( \vec{r}\right) \hat{\Phi}^{+}\left(
\vec{r}^{\prime }\right) V\left( \vec{r}-\vec{r}^{\prime }\right) \hat{\Phi}%
\left( \vec{r}\right) \hat{\Phi}\left( \vec{r}^{\prime }\right) ,
\end{eqnarray}
where $\hat{\Phi}\left( \vec{r}\right) $ and $\hat{\Phi}^{+}\left( \vec{r}%
\right) $ are the boson field operators that annihilate and create a
particle at the position $\vec{r}$, respectively, and $V\left( \vec{r}-\vec{r%
}^{\prime }\right) $ is the two-body interatomic potential \citep{Da99, rev5}. $%
V_{rot}\left( \vec{r}\right) $ is the potential associated to the rotation
of the condensate.

The use of some approximate methods can lead to a significant
simplification of the formalism. One such approach is the mean field
description of the condensate, which is based on the idea of separating out
the condensate contribution to the bosonic field operator. For a uniform gas
in a volume $V$, BEC occurs in the single particle state $\Phi _{0}=1\sqrt{V}
$, having zero momentum. The field operator can be decomposed then in the
form $\hat{\Phi}\left( \vec{r}\right) =\sqrt{N/V}+\hat{\Phi}^{\prime }\left(%
\vec{r}\right) $. By treating the operator $\hat{\Phi}^{\prime }\left( \vec{r%
}\right) $ as a small perturbation, one can develop the first order theory
for the excitations of the interacting Bose gases \citep{Da99,Ba01}.

In the general case of a non-uniform and time-dependent configuration the
field operator in the Heisenberg representation is given by $\hat{\Phi}%
\left( \vec{r},t\right) =\psi \left( \vec{r},t\right) +\hat{\Phi}^{\prime
}\left( \vec{r},t\right) $, where $\psi \left( \vec{r},t\right) $, also
called the condensate wave function, is the expectation value of the field
operator, $\psi \left( \vec{r},t\right) =\left\langle \hat{\Phi}\left( \vec{r%
},t\right) \right\rangle $. It is a classical field, and its absolute value
fixes the number density of the condensate through $\rho \left( \vec{r}%
,t\right) =\left| \psi \left( \vec{r},t\right) \right| ^{2}$. The
normalization condition is $N=\int \rho \left( \vec{r},t\right) d^{3}\vec{r}$%
, where $N$ is the total number of particles in the condensate.

The equation of motion for the condensate wave function is given by the
Heisenberg equation corresponding to the many-body Hamiltonian given by Eq.~(%
\ref{ham}),
\begin{eqnarray}
&&i\hbar \frac{\partial }{\partial t}\hat{\Phi}\left( \vec{r},t\right) =\left[
\hat{\Phi},\hat{H}\right] = \nonumber\\
&&\left[ -\frac{\hbar ^{2}}{2m}\nabla
^{2}+V_{rot}\left( \vec{r}\right) +V_{ext}\left( \vec{r}\right)\right. +\nonumber\\
&&\left.\int d\vec{r%
}^{\prime }\hat{\Phi}^{+}\left( \vec{r}^{\prime },t\right) V\left( \vec{r}%
^{\prime }-\vec{r}\right) \hat{\Phi}\left( \vec{r}^{\prime },t\right) \right]
\hat{\Phi}\left( \vec{r},t\right).  \label{gp}
\end{eqnarray}

The zeroth-order approximation to the Heisenberg
equation is obtained by replacing $\hat{\Phi}\left( \vec{r},t\right) $ with the condensate wave
function $\psi $. In the integral containing the particle-particle interaction $%
V\left( \vec{r}^{\prime }-\vec{r}\right) $ this replacement is in general a
poor approximation for short distances. However, in a dilute and cold gas,
only binary collisions at low energy are relevant, and these collisions are
characterized by a single parameter, the $s$-wave scattering length $l_a$,
independently of the details of the two-body potential. Therefore, one can
replace $V\left( \vec{r}^{\prime }-\vec{r}\right) $ with an effective
interaction $V\left( \vec{r}^{\prime }-\vec{r}\right) =\lambda \delta \left(
\vec{r}^{\prime }-\vec{r}\right) $, where the coupling constant $\lambda $
is related to the scattering length $l_a$ through $\lambda =4\pi \hbar ^{2}l_a/m$, where $m$ is the mass of the condensed particles.
With the use of the effective potential the integral in the bracket of Eq.~(%
\ref{gp}) gives $\lambda \left| \psi \left( \vec{r},t\right) \right| ^{2}$,
and the resulting equation is the Schrodinger equation with a quartic
nonlinear term \citep{rev5}. However, in order to obtain a more general
description of the Bose-Einstein condensate stars, we shall assume an
arbitrary non-linear term $g\left( \left| \psi \left( \vec{r},t\right)
\right| ^{2}\right)=g\left(\rho \right)$ \citep{Ba01}.

Therefore the generalized Gross-Pitaevskii equation describing a
gravitationally trapped rotating Bose-Einstein condensate is given by
\begin{eqnarray}\label{sch}
i\hbar \frac{\partial }{\partial t}\psi \left( \vec{r},t\right) &=&\bigg[ -%
\frac{\hbar ^{2}}{2m}\nabla ^{2}+V_{rot}\left( \vec{r}\right) +V_{ext}\left(
\vec{r}\right) +\nonumber\\
&&g^{\prime }\left( \left| \psi \left( \vec{r},t\right)
\right| ^{2}\right) \bigg] \psi \left( \vec{r},t\right) ,
\end{eqnarray}
where we denoted $g^{\prime }=dg/d\rho $.  As for $V_{ext}\left( \vec{r}%
\right) $, we assume that it is the gravitational potential $V$, $V_{ext}=V$%
, and it satisfies the Poisson equation
\begin{equation}
\nabla ^{2}V=4\pi G\rho _{m},
\end{equation}
where $\rho _{m}=m\rho =m\left| \psi \left( \vec{r},t\right) \right| ^{2}$
is the mass density inside the Bose-Einstein condensate.

\subsection{The hydrodynamical representation}

The physical properties of a Bose-Einstein condensate described by the
generalized Gross-Pitaevskii equation given by Eq.~(\ref{sch}) can be
understood much easily by using the so-called Madelung representation of the
wave function \citep{Da99}, which consist in writing $\psi $ in the form
\begin{equation}
\psi \left( \vec{r},t\right) =\sqrt{\rho \left( \vec{r},t\right) }\exp \left[
\frac{i}{\hbar }S\left( \vec{r},t\right) \right] ,
\end{equation}
where the function $S\left( \vec{r},t\right) $ has the dimensions of an
action. By substituting the above expression of  $\psi \left( \vec{r}%
,t\right) $ into Eq.~(\ref{sch}), it decouples into a system of two
differential equations for the real functions $\rho _{m}$ and $\vec{v}$,
given by
\begin{equation}
\frac{\partial \rho _{m}}{\partial t}+\nabla \cdot \left( \rho _{m}\vec{v}%
\right) =0,  \label{cont}
\end{equation}
\begin{eqnarray}
&&\rho _{m}\left[ \frac{\partial \vec{v}}{\partial t}+\left( \vec{v}\cdot
\nabla \right) \vec{v}\right] =-\nabla P\left( \frac{\rho _{m}}{m}\right)-\nonumber\\
&&\rho _{m}\nabla \left( \frac{V_{rot}}{m}\right) -
\rho _{m}\nabla \left(
\frac{V_{ext}}{m}\right) -\nabla V_{Q},  \label{euler}
\end{eqnarray}
where we have introduced the quantum potential
\begin{equation}
V_{Q}=-\frac{\hbar ^{2}}{2m}\frac{\nabla ^{2}\sqrt{\rho _{m}}}{\sqrt{\rho
_{m}}},
\end{equation}
and the velocity of the quantum fluid
\begin{equation}
\vec{v}=\frac{\nabla S}{m},
\end{equation}
respectively, and we have denoted
\begin{equation}
P\left( \frac{\rho _{m}}{m}\right) =g^{\prime }\left( \frac{\rho _{m}}{m}%
\right) \frac{\rho _{m}}{m}-g\left( \frac{\rho _{m}}{m}\right) .
\label{state}
\end{equation}

From its definition it follows that the velocity field is irrotational,
satisfying the condition $\nabla \times \vec{v}=0$. Therefore the equations
of motion of the gravitational ideal Bose-Einstein condensate take the form
of the equation of continuity and of the hydrodynamic Euler equations. The
Bose-Einstein gravitational condensate can be described as a gas whose
density and pressure are related by a barotropic equation of state %
\citep{rev6}. The explicit form of this equation depends on the form of the
non-linearity term $g$.

When the number of particles in the gravitationally bounded Bose-Einstein
condensate becomes large enough, the quantum pressure term makes a
significant contribution only near the boundary of the condensate. Hence it
is much smaller than the non-linear interaction term. Thus the quantum
stress term in the equation of motion of the condensate can be neglected.
This is the Thomas-Fermi approximation, which has been extensively used for
the study of the Bose-Einstein condensates \citep{Da99,rev5}. As the number of
particles in the condensate becomes infinite, the Thomas-Fermi approximation
becomes exact. This approximation also corresponds to the
classical limit of the theory (it corresponds to neglecting all terms with
powers of $\hbar $, or, equivalently, to the regime of strong repulsive interactions among
particles). From a mathematical point of view, the Thomas-Fermi approximation
corresponds to neglecting in the equation of motion all terms containing $%
\nabla {\rho }$ and $\nabla {S}$.

In the standard approach to the Bose-Einstein condensates, the non-linearity
term $g$ is given by
\begin{equation}
g\left( \rho _m\right) =\frac{u_{0}}{2}\left| \psi \right| ^{4}=\frac{u_{0}}{2}%
\rho _m^{2},
\end{equation}
where $u_{0}=4\pi \hbar ^{2}l_a/m$ \citep{Da99, rev5}. The corresponding equation of
state of the condensate is
\begin{equation}
P\left( \rho _{m}\right) =U_{0}\rho _{m}^{2},
\end{equation}
with
\begin{eqnarray}
U_{0}&=&\frac{2\pi \hbar ^{2}l_a}{m^{3}}= \nonumber\\
&&1.232\times10^{50}\left(\frac{m}{1\;{\rm meV}}\right)^{-3}\left(\frac{l_a}{10^9\;{\rm fm}}\right)\;{\rm cm}^5/{\rm g}\;{\rm s}^2.
\end{eqnarray}

Therefore the equation of state of the Bose-Einstein condensate with quartic
non-linearity is a polytrope with index $n=1$. However, in the case of  low dimensional systems \citet{Kolo} have shown that in many experimentally interesting cases the nonlinearity will be cubic, or even logarithmic, in $\rho _m$. The strong interaction assumption is valid only if the interaction energy per particle is much bigger than the
ground-state energy (due to the zero-point motion) per particle. This is the case for condensates in the
dilute limit below two dimensions. But as space dimensionality decreases, it becomes increasingly harder for the repulsive particles to avoid collisions. Thus the correlations between particle dominate, and the quartic nonlinearity should be replaced by a more general, power-law term \citep{Kolo}.
Hence more general models, with the non-linearity term of the form $g\left(\rho _m\right)=\alpha \rho _m^{\Gamma }$, where $\alpha =$ constant and $\Gamma =$ constant, can also be considered. In this case  the equation of state of the gravitational Bose-Einstein condensate is the standard polytropic equation of state, $P\left(\rho _m\right)=\alpha \left(\Gamma -1\right)\rho _m^{\Gamma}$, and the structure of the
static gravitationally bounded Bose-Einstein condensate is described by the Lane-Emden equation, $\left(1/\xi ^2\right)d\left(\xi^2d\theta /d\xi\right)/d\xi +\theta ^n=0$, where $n=1/\left(\Gamma -1\right)$ and $\theta $ is a dimensionless variable defined by $\rho =\rho _{cm}\theta ^n$, where $\rho _{cm}$ is the central density of the condensate. The dimensionless radial coordinate $\xi $ is defined by the relation $r=\left[(n+1)K\rho _{cm}^{1/n-1}/4\pi G\right]^{1/2}\xi $.  Hence Bose-Einstein condensate dark matter can generally be described as fluid satisfying a polytropic equation of state of index $n$.

In the following we will consider only the case of the condensate with quartic non-linearity. In this case the physical properties of the condensate are relatively well known from laboratory experiments, and its properties can be described in terms of only two free parameters, the mass $m$ of the condensate particle, and the scattering length $l_a$, respectively.

\subsection{Dark matter as a Bose-Einstein condensate}

In the case of a static
Bose-Einstein condensate, all physical quantities are independent of time.
Moreover, in the first approximation we can also neglect the rotation of the
condensate, by taking $V_{rot}=0$. Therefore the equations describing the
static Bose-Einstein condensate in a gravitational field with potential $V$
take the form
\begin{equation}
\nabla P\left( \frac{\rho _{BE}}{m}\right) =-\rho _{BE}\nabla \left( \frac{V}{m%
}\right) ,
\end{equation}
\begin{equation}
\nabla ^{2}V=4\pi G\rho _{BE}.
\end{equation}

These equations must be integrated together with the equation of state $%
P=P\left( \rho _{BE}\right) =U_{0}\rho _{BE}^{2}$,  and some appropriately chosen boundary conditions. The density
distribution $\rho _{BE}$ of the static gravitationally bounded single
component dark matter Bose-Einstein condensate is given by  \citep{BoHa07}%
\begin{equation}
\rho _{BE}\left( r\right) =\rho _{BE}^{(c)}\frac{\sin kr}{kr}
\end{equation}
where $k=\sqrt{Gm^{3}/\hbar ^{2}l_a}$ and $\rho _{BE}^{(c)}$ is the central
density of the condensate, $\rho _{BE}^{(c)}=\rho _{BE}(0)$. The mass
profile $m_{BE}(r)=4\pi \int_{0}^{r}\rho _{BE}(r)r^{2}dr$ of the
Bose-Einstein condensate galactic halo is
\begin{equation}
m_{BE}\left( r\right) =\frac{4\pi \rho _{BE}^{(c)}}{k^{2}}r\left( \frac{\sin
kr}{kr}-\cos kr\right) ,
\end{equation}
with a boundary radius $R_{BE}$. At the boundary of the dark matter
distribution $\rho _{BE}(R_{BE})=0$, giving the condition $kR_{BE}=\pi $,
which fixes the radius of the condensate dark matter halo as $R_{BE}=\pi
\sqrt{\hbar ^{2}l_a/Gm^{3}}$. The tangential velocity of a test particle
moving in the condensed dark halo can be represented as \citep{BoHa07}
\begin{equation}
V_{BE}^{2}\left(
r\right) =\frac{Gm_{BE}(r)}{r}= \frac{4\pi G\rho _{BE}^{(c)}}{k^{2}} \left(
\frac{\sin kr}{kr}-\cos kr\right)  .
\end{equation}
The mass of the particle in the
condensate can be obtained from the radius of the dark matter halo in the
form \citep{BoHa07}
\begin{eqnarray}
m &=&\left( \frac{\pi ^{2}\hbar ^{2}l_a}{GR_{BE}^{2}}\right) ^{1/3}\approx \nonumber\\
&&2.58\times 10^{-30}\left[ l_a\left( {\rm cm}\right) \right] ^{1/3}\left[
R_{BE}\;{\rm (kpc)}\right] ^{-2/3}\;{\rm g}\approx  \nonumber \\
&&6.73\times 10^{-2}\left[ l_a\left( {\rm fm}\right) \right] ^{1/3}%
\left[ R_{BE}\;{\rm (kpc)}\right] ^{-2/3}\;{\rm eV}.
\end{eqnarray}

From this equation it follows that $m$ is of the order of eV. For $l_a\approx
1 $ fm and $R_{BE}\approx 10$ kpc, the mass is of the order of $m\approx 14$
meV. For values of $l_a$ of the order of $l_a\approx 10^{6}$ fm, corresponding
to the values of $l_a$ observed in terrestrial laboratory experiments, $%
m\approx 1.44$ eV. These values are perfectly consistent with the limit $%
m<1.87$ eV obtained for the mass of the condensate particle from
cosmological considerations \citep{Fuk08}.

\section{Post-Newtonian hydrodynamics of the Bose-Einstein condensates}\label{sect3}

In order to study gravitational effects on the evolution of
Bose-Einstein condensate dark halos a full general relativistic treatment is
needed. The equations of motion of the condensate are obtained from the
conservation of the energy-momentum tensor, $T_{\nu ;\mu }^{\mu }=0$, with $%
;$ denoting the covariant derivative with respect to the metric $g_{\mu \nu }$%
, and
\begin{equation}
T_{\nu }^{\mu }=\left( \rho _{m}c^{2}+P\right) u_{\nu }u^{\mu }-P\delta
_{\nu }^{\mu },
\end{equation}
where $u^{\mu }$ is the four-velocity of the fluid, satisfying the
conditions $u_{\mu }u^{\mu }=1$ and $u_{;\nu }^{\mu }u_{\mu }=0$, respectively. By taking
the covariant divergence of $T_{\nu }^{\mu }$ we obtain the equation
\begin{eqnarray}
&&\left( \rho _{m}c^{2}+P\right) _{,\mu }u_{\nu }u^{\mu }+\left( \rho
_{m}c^{2}+P\right) u_{\nu ;\mu }u^{\mu }+\nonumber\\
&&\left( \rho _{m}c^{2}+P\right)
u_{\nu }u_{;\mu }^{\mu }=P_{,\nu },  \label{cons}
\end{eqnarray}
where a comma denotes the ordinary derivative with respect to the coordinate
$x^{\mu }$. By contracting Eq.~(\ref{cons}) with $u^{\nu }$, we obtain
\begin{equation}
\left( \rho _{m}c^{2}+P\right) _{,\mu }u^{\mu }+\left( \rho
_{m}c^{2}+P\right) u_{;\mu }^{\mu }=P_{,\nu }u^{\nu }.  \label{contr1}
\end{equation}

In the Newtonian limit of small condensate velocities the four-velocity is
given by $u^{\mu }=\left( 1,\vec{v}/c\right) $, where $\vec{v}$ is the
three-velocity of the condensate. The four-divergence of the
four-velocity is given by $u_{;\mu }^{\mu }=\left( 1/\sqrt{-g}\right)
\partial \left( \sqrt{-g}u^{\mu }\right) /\partial x^{\mu }$, where $\left(-g\right)$ is the
determinant of the metric tensor. In the Newtonian limit we assume that $%
\sqrt{-g}\rightarrow 1$, that is, the deviations from the Minkowski type
geometry are small. Under these assumptions, from Eq.~(\ref{contr1}) we
obtain the equation of continuity of the Bose-Einstein condensate as
\begin{equation}
\left( \frac{\partial \rho _{m}}{\partial t}\right) _{r}+\nabla _{r}\cdot
\left( \rho _{m}\vec{v}\right) +\frac{P}{c^{2}}\nabla _{r}\cdot \vec{v}=0,
\label{cont1}
\end{equation}
where all differential operations are considered with respect to the
physical coordinate $\vec{r}$. By contracting Eq.~(\ref{cons}) with the
projection operator $h_{\alpha }^{\nu }=\delta _{\alpha }^{\nu }-u_{\alpha
}u^{\nu }$, with the property $h_{\alpha }^{\nu }u_{\nu }\equiv 0$, we
obtain the relativistic Euler equation of motion as
\begin{equation}
\left( \rho _{m}c^{2}+P\right) u_{\alpha ;\mu }u^{\mu }=P_{,\alpha }-P_{,\nu
}u^{\nu }u_{\alpha }.
\end{equation}

In the Newtonian approximation the generalized Euler equation of motion
becomes
\begin{equation}
\left( \frac{\partial \vec{v}}{\partial t}\right) _{r}+\left( \vec{v}\cdot
\nabla _{r}\right) \vec{v}=-\nabla _{r}V-\frac{c^{2}\nabla _{r}P+\dot{P}\vec{%
v}}{\rho _{m}c^{2}+P}.  \label{euler1}
\end{equation}

The gravitational potential $V$ satisfies the generalized Poisson equation,
\begin{equation}
\nabla _{r}^{2}V=4\pi G\left( \rho _{m}+3\frac{P}{c^{2}}\right) .
\label{poi1}
\end{equation}

Eqs.~(\ref{cont1}), (\ref{euler1}) and (\ref{poi1}) represent the basic
equations describing the dynamics of a gravitationally bounded Bose-Einstein
condensate in the first Post-Newtonian approximation \citep{McCrea51, Har64, Lima97, Reis03, Ab07, Pace10}.

\section{Cosmological dynamics of Bose-Einstein condensates}\label{sect4}

The Bose-Einstein condensation takes place when particles (bosons) become
correlated with each other. This happens when their wavelengths overlap,
that is, the thermal wavelength $\lambda _{T}=\sqrt{2\pi \hbar ^{2}/mk_{B}T}$
is greater than the mean inter-particles distance $a$, $\lambda _{T}>a$. The
critical temperature for the condensation to take place is $T_{cr}<2\pi \hbar
^{2}n^{2/3}/mk_{B}$ \citep{Da99}. On the other hand, cosmic evolution has the same temperature dependence, since
in an adiabatic expansion process the density of a matter dominated Universe evolves as $\rho \propto T^{3/2}$ \citep{Fuk09}. Therefore, if the boson temperature is equal, for example, to the radiation temperature at $z = 1000$,
 the critical temperature for the Bose-Einstein condensation is at present $T_{cr} = 0.0027K$ \citep{Fuk09}. Since the matter temperature $T_m$ varies as $T_m\propto  a^{-2}$, where $a$ is the scale factor of the Universe, it follows that during an adiabatic evolution the ratio of the photon temperature $T_{\gamma }$ and of the matter temperature evolves as $T_{\gamma }/T_m\propto a$. Using for the present day energy density of the Universe
the value $\rho _{cr}= 9.44 \times 10^{-30}$ g/cm$^3$, BEC takes place provided that the boson mass satisfies the restriction $m < 1.87$ eV \citep{Fuk08}.
Thus, once the
temperature $T_{cr}$ of the boson is less than the critical temperature,  BEC can always take place at some
moment during the cosmological evolution of the Universe. On the other hand,  we expect that the Universe is always under critical temperature, if it is at the present time \citep{Fuk09}. Another cosmological bound on the mass of the condensate particle can be obtained as $m<2.696\left(g_d/g\right)\left(T_d/T_{cr}\right)^3$ eV \citep{Bo}, where $g$ is the number of internal degrees of freedom of the particle before decoupling, $g_d$ is the number of internal degrees of freedom of the particle at the decoupling, and $T_d$ is the decoupling temperature. In the Bose condensed case $T_d/T_c < 1$, and it follows that the BEC particle should be light, unless it decouples very early on, at high temperature and with a large $g_d$. Therefore,  depending on the relation between the critical  and the decoupling
temperatures, in order for a BEC light relic to act as cold dark matter,  the decoupling scale must be higher than the electroweak scale \citep{Bo}.


The set of equations Eqs.~(\ref{cont1}), (\ref{euler1}) and (\ref{poi1})
admits a homogeneous and isotropic cosmological background solution with $%
\rho _{m}=\rho _{b}(t)$ and $P=P_{b}\left( t\right) $. In this case the
fluid's velocity is given by
\begin{equation}
\vec{v}_{b}=\frac{\dot{a}}{a}\vec{r},
\end{equation}
and the evolution of the scale factor $a$ is determined by the Friedmann
equations,
\begin{equation}
3\frac{\dot{a}^{2}}{a^{2}}=3H^{2}=8\pi G\rho _{b},  \label{fried1}
\end{equation}
and
\begin{equation}
\frac{\ddot{a}}{a}=-\frac{4\pi G}{3}\left( \rho _{b}+\frac{P_{b}}{c^{2}}%
\right) ,  \label{fried2}
\end{equation}
respectively, where we have denoted $H=\dot{a}/a$. The continuity equation
Eq.~(\ref{cont1}) reduces to
\begin{equation}
\frac{d\rho _{b}}{dt}+3H\left( \rho _{b}+\frac{P_{b}}{c^{2}}\right) =0.
\label{fried3}
\end{equation}

In the case of the Bose-Einstein condensates the equation of state is given
by $P_{b}=U_{0}\rho _{b}^{2}$, and Eq.~(\ref{fried3}) can be integrated
immediately to obtain
\begin{equation}
\rho _{b}\left( a\right) =\frac{C}{a^{3}-CU_{0}/c^{2}},
\end{equation}
where $C$ is an arbitrary constant of integration. By assuming that the present-day density of the Bose-Einstein
condensate, $\rho _{m,0}$,  is obtained for a value $a=a_0$ of the scale factor, we obtain $C=\rho _{m,0}a_0^3/\left( 1+\rho
_{m,0}U_{0}/c^{2}\right) $, and the background cosmological density of the
condensate can be written as
\begin{equation}
\rho _{b}\left( a\right) =\frac{c^{2}}{U_{0}}\frac{\rho _{0}}{\left(a/a_0\right)^{3}-\rho _{0}%
},
\end{equation}
where we have denoted
\begin{equation}
\rho _{0}=\frac{\rho _{m,0}U_{0}/c^{2}}{
1+\rho _{m,0}U_{0}/c^{2}} .
\end{equation}

The energy density of the Bose-Einstein condensate diverges as $a\rightarrow
\rho _{0}^{1/3}$. The equation determining the time
evolution of the scale factor is given by
\begin{equation}\label{scale}
\frac{da}{dt}=H_{0}\sqrt{\Omega_{BE}}%
\frac{a}{\sqrt{\left(a/a_0\right)^{3}-\rho _{0}}},
\end{equation}
where we have denoted
\begin{equation}
\Omega _{BE}=\frac{\Omega _{BE,0}}{1+\rho _{m,0}U_{0}/c^{2}}=\frac{\Omega _{BE,0}}{1+\Omega _{BE,0}\rho _{cr,0}U_{0}/c^{2}},
\end{equation}
where $\Omega _{BE,0}=\rho _{m,0}/\rho _{cr,0}$ is the present day density
parameter of the Bose-Einstein condensate, $\rho _{cr,0}=3H_{0}^{2}/8\pi G$
is the present day critical density of the Universe, and $H_{0}$ is the
present day value of the Hubble parameter, respectively.

Eq.~(\ref{scale}) can be integrated immediately to give the time evolution
of the scale factor of the Bose-Einstein condensate as
\begin{eqnarray}
&&\sqrt{\Omega _{BE}}H_{0}\left( t-C_1\right) =\frac{2}{3}\sqrt{\left(\frac{a}{a_0}\right)^{3}-\rho
_{0}}-\nonumber\\
&&\frac{2}{3}\sqrt{\rho _{0}}\arctan \left[ \sqrt{\frac{\left(a/a_0\right)^{3}-\rho _{0}}{%
\rho _{0}}}\right] ,
\end{eqnarray}%
where $C_1$ is an arbitrary constant of integration. The constant $C_1$ can be determined from the condition $t=0$ when $\left(a/a_0\right)^{3}=\rho
_{0}$, thus obtaining $C_1=0$. Therefore the time evolution of the scale factor is described by the equation
\begin{eqnarray}
\frac{t}{t_H}&=&\frac{2}{3\sqrt{\Omega _{BE}}}\left\{\sqrt{\left(a/a_0\right)^{3}-\rho
_{0}}\right.-\nonumber\\
&&\left.\sqrt{\rho _{0}}\arctan \left[ \sqrt{\frac{\left(a/a_0\right)^{3}-\rho _{0}}{%
\rho _{0}}}\right]\right\},
\end{eqnarray}
where we denoted $t_H=1/H_0$.

In the case of the
standard dark matter models, dark matter is assumed to be a pressureless
fluid, and the background cosmological evolution is described by the
Einstein-de Sitter model, with the scale factor given by $a/a_0=\left(9\Omega _{DM,0}/4\right)^{1/3} \left(t/t_H\right)^{2/3}$, where
$\Omega _{DM,0}$ is the present day density parameter of the dark matter, and we have assumed that $a(0)=0$. In
the following for the Hubble constant we adopt the value $%
H_{0}=70\;{\rm km}/{\rm s}/{\rm Mpc}=2.273\times 10^{-18}\;{\rm s}^{-1}$ \citep{Hin09}, giving for the critical
density a value of $\rho _{c,0}=9.248\times 10^{-30}\;{\rm g}/{\rm cm}^{3}$. The constant $\rho _{0}$ can be represented as
\begin{equation}
\rho _{0}=\frac{1.266\times\Omega _{BE,0} \times \left( m/1\;{\rm meV}\right)
^{-3}\times \left( l_{a}/10^{9}\;{\rm fm}\right) }{1+1.266\times\Omega _{BE,0}\times
\left( m/1\;{\rm meV}\right) ^{-3}\times \left( l_{a}/10^{9}\;{\rm fm}\right) },
\end{equation}
while for $\Omega _{BE}$ we obtain
\begin{equation}
\Omega _{BE}=\frac{\Omega _{BE,0}}{1+1.266\times\Omega _{BE,0}\times \left(
m/1\;{\rm meV}\right) ^{-3}\times \left( l_{a}/10^{9}\;{\rm fm}\right) }.
\end{equation}
The two parameters $\rho_0$ and $\Omega _{BE}$, describing the global properties of the condensate, are related by the relation $\Omega _{BE}=\Omega _{DM,0}\times\left(1-\rho _0\right)$.

\begin{figure}
\centering
\includegraphics[width=8.15cm]{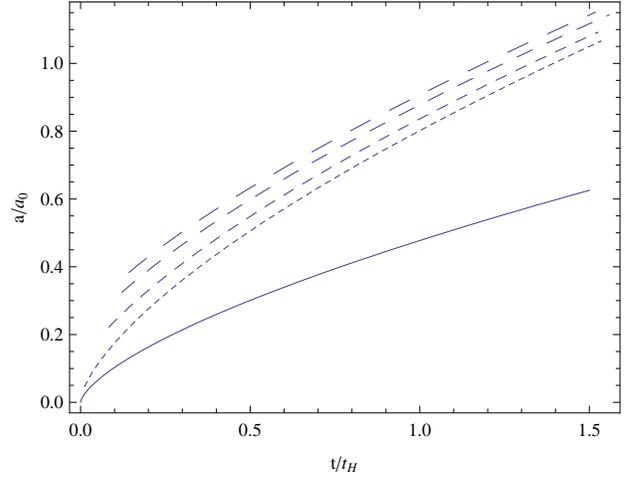}
\caption{Time evolution of the scale factor of the pressureless dark matter, described by the Einstein-de Sitter metric (solid curve), and of the Bose-Einstein condensate dark matter, for different values of $\rho _0$: $\rho _0=10^{-6}$ (dotted curve), $\rho _0=10^{-3}$ (dashed curve), $\rho _0=0.005$ (long-dashed curve), and $\rho _0=0.01$ (ultra-long dashed curve).}
\label{fig1}
\end{figure}

By assuming that the entire existing dark matter is in the form of a Bose-Einstein condensate, it follows that $\Omega _{BE,0}\approx \Omega _{DM,0}\approx 0.228$ \citep{Hin09}. By assuming that $m=1$ meV and $l_a=10^{10}$ fm, we obtain $\rho _0=0.7426$, while $\Omega _{BE}=0.05866$. For these values of the physical parameters of the condensate the energy density of the dark matter diverges for $a/a_0\rightarrow0.9056$. The time evolution of the scale factor $a$ for the Bose-Einstein condensate dark matter, for different values of the parameters $m$ and $l_a$, and for the pressureless dark matter, are represented in Fig.~\ref{fig1}, respectively. The cosmological dynamics of the condensate dark matter shows significant differences as compared to the standard pressureless dark matter model, with the condensate expanding much faster than the cosmological fluid of the standard $\Lambda $CDM model, with the speed of expansion increasing with increasing $\rho _0$.

In the case of a Universe filled with dark energy, radiation, baryonic
matter with negligible pressure, and Bose-Einstein condensed dark matter,
respectively, the time evolution of the scale factor is given by
\begin{equation}
\frac{1}{a}\frac{da}{dt}=H_{0}\sqrt{\frac{\Omega _{B,0}}{\left(
a/a_{0}\right) ^{3}}+\frac{\Omega _{rad,0}}{\left( a/a_{0}\right) ^{4}}+%
\frac{\Omega _{BE}}{\left( a/a_{0}\right) ^{3}-\rho _{0}}+\Omega _{\Lambda }},
\end{equation}
where $\Omega _{B,0}$, $\Omega _{rad,0}$, and $\Omega _{\Lambda }$ are the
present day values of the density parameters of the baryonic matter,
radiation, and dark energy, respectively. For $\Omega _{B,0}$, $\Omega
_{rad,0}$, and $\Omega _{\Lambda }$ we adopt the numerical values $\Omega
_{B,0}=0.0456$, $\Omega _{rad,0}=8.24\times 10^{-5}$ and $\Omega _{\Lambda
}=0.726$ \citep{Hin09}. In the case of the standard $\Lambda $CDM cosmological, model the
Friedmann equation describing the evolution of the Universe containing baryons,
pressureless dark matter, radiation, and dark energy, is given by
\begin{equation}
\frac{1}{a}\frac{da}{dt}=H_{0}\sqrt{\frac{\Omega _{B,0}+\Omega _{DM,0}}{%
\left( a/a_{0}\right) ^{3}}+\frac{\Omega _{rad,0}}{\left( a/a_{0}\right) ^{4}%
}+\Omega _{\Lambda }}.
\end{equation}

\begin{figure}
\centering
\includegraphics[width=8.15cm]{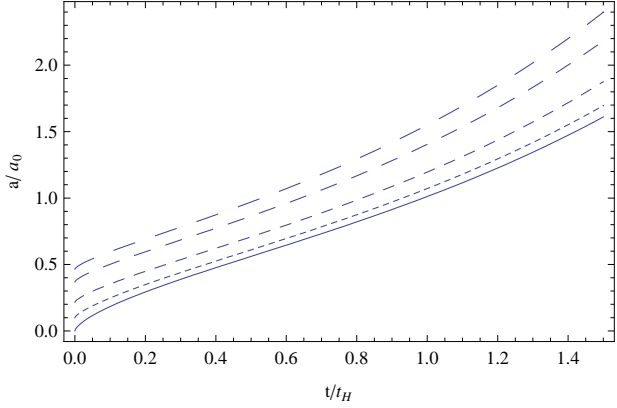}
\caption{Time evolution of the scale factor of a Universe filled with dark energy, baryonic matter and with Bose-Einstein condensate dark matter, respectively, for different values of $\rho _0$: $\rho _0=10^{-3}$ (dotted curve), $\rho _0=10^{-2}$ (dashed curve), $\rho _0=0.05$ (long-dashed curve), and $\rho _0=0.1$ (ultra-long dashed curve). The solid curve represents the time evolution of the scale factor in the standard $\Lambda $CDM model, in which the dark matter is pressureless.}
\label{fig1_1}
\end{figure}

The time evolutions of the scale factors for Universes containing BEC dark
matter and standard pressureless dark matter are represented, for different
values of the BEC parameter $\rho _{0}$,  in Fig.~\ref{fig1_1}. The presence of
the condensate dark matter changes the global cosmological
dynamics of the Universe, and the magnitude of the changes increases with the increase of the BEC parameter $\rho _0$.

\section{Cosmological perturbations of an expanding Bose-Einstein condensate}\label{sect5}

In the gravitationally bounded Bose-Einstein condensate we assume small perturbations of the
physical quantities around the homogeneous background of the form
\begin{equation}
\rho _{m}\left( \vec{r},t\right) =\rho _{b}\left( t\right) +\delta \rho
\left( \vec{r},t\right) ,  \label{pert1}
\end{equation}
\begin{equation}
P\left( \vec{r},t\right) =P_{b}\left( t\right) +\delta P\left( \vec{r}%
,t\right) ,  \label{pert2}
\end{equation}
\begin{equation}
V\left( \vec{r},t\right) =V_{b}+\varphi \left( \vec{r},t\right) ,
\label{pert3}
\end{equation}
\begin{equation}
\vec{v}\left( \vec{r},t\right) =\vec{v}_{b}+\vec{u}\left( \vec{r},t\right) .
\label{pert4}
\end{equation}

In Eqs.~(\ref{pert1})-(\ref{pert4}) the index $b$ denotes the background
quantities.  Substituting these equations into the continuity equation Eq.~(%
\ref{cont1}) we obtain
\begin{equation}
\left( \frac{\partial \delta \rho }{\partial t}\right) _{r}+\left( \rho
_{b}+\frac{P_{b}}{c^{2}}\right) \nabla _{r}\cdot \vec{u}+\nabla _{r}\cdot \left(
\delta \rho \vec{v}_{b}\right) +\frac{\delta P}{c^2}\nabla _{r}\cdot \vec{v}_{b}=0.
\end{equation}

The variation of the equation of motion Eq.~(\ref{euler1}) gives
\begin{eqnarray}\label{pb}
&&\left( \frac{\partial \vec{u}}{\partial t}\right) _{r}+\left( \vec{v}%
_{b}\cdot \nabla _{r}\right) \vec{u}+\left( \vec{u}\cdot \nabla _{r}\right)
\vec{v}_{b}= \nonumber\\
&&-\nabla _{r}\varphi -\frac{\nabla _{r}\delta P+\dot{P}_{b}\vec{u}%
/c^{2}}{\rho _{b}+P_{b}/c^{2}}.
\end{eqnarray}

The Poisson equations for the perturbation of the gravitational potential,
obtained by perturbing Eq.~(\ref{poi1}), can be written as
\begin{equation}
\nabla _{r}^{2}\varphi =4\pi G\left( \delta \rho +3\frac{\delta P}{c^{2}}%
\right) .
\end{equation}

In order to describe the cosmological evolution we make a change to the
comoving coordinate system, so that $\vec{r}=a\vec{q}$, $\nabla _{q}=\nabla
=a\nabla _{r}$, and
\begin{equation}
\left( \frac{\partial }{\partial t}\right) _{q}=\frac{\partial }{\partial t}%
=\left( \frac{\partial }{\partial t}\right) _{r}+\frac{\dot{a}}{a}\left(
\vec{q}\cdot \nabla _{q}\right) ,
\end{equation}
respectively. To simplify the notation we define the parameters $w=P_{b}/\rho _{b}c^{2}$ and
$c_{eff}^{2}=\delta P/\delta \rho $, respectively, which generally are functions of the
time only. We also introduce the density contrast as $\delta =\delta \rho
/\rho _{b}$. The time derivative of the background pressure is related to the speed of
sound of the background condensate  $c_{s}^{2}=\partial P_{b}/\partial \rho _{b}$ by the relation
\begin{equation}
\dot{P}
_{b}=-3\left( \frac{\dot{a}}{a}\right) c_{s}^{2}\rho _{b}(1+w).
\end{equation}
Therefore the perturbation equations Eqs.~(\ref{pert2}) - (\ref
{pert4}) can be written as
\begin{equation}
\dot{\delta}+3H\left( \frac{c_{eff}^{2}}{c^{2}}-w\right) \delta +\frac{1+w}{a%
}\nabla \cdot \vec{u}=0,  \label{fin1}
\end{equation}
\begin{equation}
\frac{d\vec{u}}{dt}+\left(1-3\frac{c_s^2}{c^2}\right)\frac{\dot{a}}{a}\vec{u}+\frac{1}{a}\nabla \varphi +%
\frac{c_{eff}^{2}}{c^{2}}\frac{1}{a}\frac{1}{1+w}\nabla \cdot \delta =0,
\label{fin2}
\end{equation}
\begin{equation}
\nabla ^{2}\varphi =4\pi Ga^{2}\rho _{b}\left( 1+3\frac{c_{eff}^{2}}{c^{2}}%
\right) \delta .  \label{fin3}
\end{equation}

In the following we denote
\begin{equation}
\alpha _{eff}=\frac{c_{eff}^{2}}{c^{2}}-w,
\end{equation}
and
\begin{equation}
\alpha _{s}=\frac{c_{s}^{2}}{c^{2}}-w,
\end{equation}
respectively. $\alpha _s$ is related to the time derivative of $w$ by the relation
\begin{equation}
\frac{\dot{w}}{1+w}=-3H\alpha _s.
\end{equation}

By taking the time derivative of Eq.~(\ref{fin1}), the divergence of Eq.~(\ref{fin2}), by eliminating $\nabla \cdot \vec{u}$ by using the perturbed equation of continuity, and with the use of Eq.~(\ref{fin3}), we obtain the equation giving the evolution of the density contrast as
\begin{eqnarray}\label{pertfin}
&&\ddot{\delta}+3H\left( \frac{c_{eff}^{2}}{c^{2}}-2w+\frac{2}{3}\right)
\dot{\delta}+\frac{3}{2}H^{2}\times \nonumber\\\
&&\left[ 9w^{2}-2w-2\left( 1+6w\right) \frac{c_{eff}^{2}}{c^{2}%
}+\frac{2}{H}\frac{d}{dt}\alpha _{eff}-1\right]  \delta=\nonumber\\
&&\frac{c_{eff}^{2}}{c^{2}}\frac{1}{a^{2}}\Delta \delta .
\end{eqnarray}

Changing the independent variable from the time $t$ to the scale factor $a$ using the relations
$\partial /\partial t=aH(a)\partial /\partial a$ and $\partial ^{2}$
$/\partial t^{2}=a^{2}H^{2}\partial ^{2}/\partial a^{2}-\left[ (1+3w)aH^{2}/2%
\right] \partial /\partial a$, we obtain the evolution equation in the form
\begin{eqnarray}\label{pertfin1}
&&a^{2}\frac{\partial ^{2}\delta }{\partial a^{2}}+3a\left( \frac{c_{eff}^{2}%
}{c^{2}}-\frac{5}{2}w+\frac{1}{2}\right) \frac{\partial \delta }{%
\partial a}+ \frac{3}{2}\times\nonumber\\
&&\left[ 9w^{2}-2w-2\left( 1+6w\right) \frac{c_{eff}^{2}}{c^{2}%
}+2a\frac{d}{da}\alpha _{eff}-1\right]  \delta =\nonumber\\
&&\frac{c_{eff}^{2}}{c^{2}}\frac{1}{a^{2}H^{2}}\Delta \delta .
\end{eqnarray}

Eq.~(\ref{pertfin}) is different from the perturbation equation obtained in the Newtonian cosmology with pressure
by \citet{Lima97}, \citet{Reis03}, and \citet{Ab07}, respectively. The reason is that we have included in our
analysis the term $\dot{P}_{b}\vec{u}/c^{2}$, which was neglected in the previous studies. This term generates the new term $-\left(3c_s^2/c^2\right)H\vec{u}$ in the left hand side of Eq.~(\ref{fin2}), which modifies the final perturbation equation. On the other hand, in the present approach the term $\delta P\left(\vec{r},t\right)/c^2$ was neglected.

In order to numerically integrate Eq.~(\ref{pertfin}) or Eq.~(\ref{pertfin1}) we have to chose some physically appropriate initial conditions. In the current standard model for structure formation in the Universe, it is supposed that quantum fluctuations were generated during an initial period of inflation. These fluctuations inflated up to super-horizon scales, producing a near scale-invariant, and near Gaussian, set of primordial potential fluctuations. At the end of inflation, the Universe is
reheated, and particles and radiation are produced. In this hot early phase, cold thermal relics (dark matter) are also formed \citep{PeRa031, PeRa032}. Dark matter particles interact gravitationally, and possibly through the weak interaction. Therefore in order to obtain some physically realistic initial conditions for the state of the Universe during large scale structure formation, one needs to evolve cosmological perturbations, starting from initial conditions, deep inside the radiation epoch, and far outside the Hubble radius. Initial conditions for photons, neutrinos, cold dark matter and baryons have been obtained, in the framework of the standard $\Lambda $CDM cosmological model,  in both the synchronous and Newtonian gauges, by \citet{Ma}. In the conventional method, the power spectrum of the matter fluctuations in the Universe is computed by numerically solving the Boltzmann equation. The power spectrum is usually obtained in the linear theory, and then extrapolated to the present epoch. In the standard cosmology, in first-order Eulerian perturbation theory, all modes evolve independently, and the power spectrum can be scaled back to the initial epoch via the growth function \citep{Ma}. Moreover, the effect of the dark matter pressure is generally ignored in the conventional methods of generating initial conditions. On the other hand, the exact moment in the history of the Universe when the Bose-Einstein condensation occurred is not known. That's why obtaining the rigorous and physically well motivated initial conditions for the density contrast $\delta $ and for its derivative for BEC dark matter requires the full investigation of the cosmological dynamics from the reheating era, by taking into account the dark matter condensation and pressure effects.

\section{Cosmological evolution of small perturbations in a Bose-Einstein condensate}\label{sect6}

By taking into account the equation of state of the Bose-Einstein
condensate, we immediately obtain $w=P_{b}/\rho _{b}c^{2}=\left(
U_{0}/c^{2}\right) \rho _{b}=\rho _{0}/\left[ \left( a/a_{0}\right)
^{3}-\rho _{0}\right] $, and $c_{s}^{2}/c^{2}=c_{eff}^{2}/c^{2}=2w$,
respectively. The conditions $c_{s}^{2}/c^{2}\leq 1$ and $%
c_{eff}^{2}/c^{2}\leq 1$ imposes the constraint $\left( a/a_{0}\right)
^{3}\geq 3\rho _{0}$, and in the following we will consider that the model
considered in the present paper is valid only for this range of values of
the scale factor. Hence for the time evolution of the linear density
perturbations of the Bose-Einstein condensate dark matter we obtain
successively
\begin{eqnarray}
&&\alpha ^{2}\frac{d^{2}\delta }{d\alpha ^{2}}+\frac{3}{2}\alpha \left(
1-w\right) \frac{d\delta }{d\alpha }-\nonumber\\
&&\frac{3}{2}\left( 1+6w+15w^{2}-2\alpha
\frac{dw}{d\alpha }\right) \delta =0,
\end{eqnarray}
and
\begin{eqnarray}
&&\alpha ^{2}\frac{d^{2}\delta }{d\alpha ^{2}}+\frac{3}{2}\alpha \left( 1-%
\frac{\rho _{0}}{\alpha ^{3}-\rho _{0}}\right) \frac{d\delta }{d\alpha }-\frac{3}{2}\times\nonumber\\
&&\left[ 1+\frac{6\rho _{0}}{\alpha ^{3}-\rho _{0}}+\frac{15\rho
_{0}^{2}}{\left( \alpha ^{3}-\rho _{0}\right) ^{2}}+\frac{6\rho _{0}\alpha
^{3}}{\left( \alpha ^{3}-\rho _{0}\right) ^{2}}\right] \delta =0,
\label{pertnum}
\end{eqnarray}
respectively, where we have denoted $\alpha =a/a_{0}$. In the limit of large
$\alpha $, when $\left( a/a_{0}\right) ^{3}>>\rho _{0}$, $w\rightarrow 0$,
and Eq.~(\ref{pertnum}) becomes
\begin{equation}
\alpha ^{2}\frac{d^{2}\delta }{d\alpha ^{2}}+\frac{3}{2}\alpha \frac{d\delta
}{d\alpha }-\frac{3}{2}\delta =0,
\end{equation}
with the solution
\begin{equation}
\delta \left( a\right) \approx C_{1}\left( \frac{a}{a_{0}}\right)
+C_{2}\left( \frac{a}{a_{0}}\right) ^{-3/2},
\end{equation}
where $C_{1}$ and $C_{2}$ are arbitrary constants of integration. Hence in
the limit of a pressureless fluid we recover the standard general
relativistic result. In the limit of small $\alpha $, so that $\alpha
^{3}\rightarrow 3\rho _{0}$, we can approximate Eq.~(\ref{pertnum}) as
\begin{equation}
\alpha ^{2}\frac{d^{2}\delta }{d\alpha ^{2}}+\frac{3}{4}\alpha \frac{d\delta
}{d\alpha }-\frac{147}{8}\delta =0,
\end{equation}
with the solution
\begin{equation}
\delta \left( a\right) \approx C_{1}^{^{\prime }}\left( \frac{a}{a_{0}}%
\right) ^{\left( 1+\sqrt{1177}\right) /8}+C_{2}^{^{\prime }}\left( \frac{a}{%
a_{0}}\right) ^{\left( 1-\sqrt{1177}\right) /8},
\end{equation}
where $C_{1}^{^{\prime }}$ and $C_{2}^{^{\prime }}$ are two arbitrary
constants of integration.

By introducing  $w=\rho _0/\left(\alpha ^3-\rho_0\right)$ as a new independent variable, Eq.~(\ref{pertnum}) can be written as
\begin{eqnarray}\label{eqw}
&&3w^{2}\left( 1+w\right) ^{2}\frac{d^{2}\delta }{dw^{2}}+\frac{5}{2}w\left( 1+w\right)
\left( 1+3w\right) \frac{d\delta }{dw}-\nonumber\\
&&\frac{1}{2}\left(1+12w+21w^{2}\right) \delta =0.
\end{eqnarray}
By representing the density contrast $\delta $ as $\delta (w)=w^{-1/3}(1+w)^{-5/3}u(w)$, it follows that the new function $u(w)$ satisfies the equation
\begin{equation}
6w\left( 1+w\right) \frac{d^{2}u}{dw^{2}}+(1-9w)\frac{du}{dw}-15u=0.
\end{equation}
Therefore the general solution of Eq.~(\ref{eqw}) can be obtained as
\begin{eqnarray}
&&\delta (w)= w^{-1/3}(1+w)^{-5/3}\times \nonumber\\
&&\left[C_1^{\prime} \;
   _2F_1\left(-\frac{5+\sqrt{65}}{4} ,\frac{-5+\sqrt{65}}{4}
   ;\frac{1}{6};-w\right)\right. +\nonumber\\
&& \left.C_2^{\prime} w^{5/6}\;
   _2F_1\left(-\frac{5}{12}-\frac{\sqrt{65}}{4}
   ,-\frac{5}{12}+\frac{\sqrt{65}}{4}
   ;\frac{11}{6};-w\right)\right],\nonumber\\
\end{eqnarray}
where $_2F_1\left(a,b;c;z\right)$ is the hypergeometric function, $_{2}F_{1}\left( a,b;c;z\right) =\sum_{k=0}^{\infty }\left( a\right)
_{k}(b)_{k}z^{k}/(c)_{k}k!$, $|z|<1$, and $C_1^{\prime}$ and $C_2^{\prime}$ are arbitrary constants of integration. The constants of integration can be determined from the initial conditions. When $a^3=3\rho _0$, $w=1/2$, and $u(1/2)=\left(3^{5/3}/4\right)\delta _i$ and $u'\left(1/2\right)=\left(3^{5/3}/4\right)\left[\delta '_i +16\delta _i/9\right]$, where we have denoted $\delta _i=\delta (1/2)$ and $\delta '_i=\delta '(1/2)$, respectively. In the limit of large $a$, $w\rightarrow 0$, and the density contrast can be approximated as
\begin{eqnarray}
&&\delta (w)\approx0.00231255 \left(347.503 \delta _i-250.317 \delta
   '_i\right)w^{-1/3}+\nonumber\\
   &&0.00925 \left[1533.98
   \left(0.888 \delta _i+\frac{\delta
   '_i}{2}\right)-2366.51 \delta _i\right]
   \sqrt{w}-\nonumber\\
   &&0.0308341 \left(250.317 \delta
   '_i-347.503 \delta _i\right) w^{2/3}+0.00925\times\nonumber\\
  && \bigg\{-5019.88 \delta _i-1.66667 \times \nonumber\\
  &&\left[1533.98
   \left(0.888 \delta _i +\frac{\delta
   '_i}{2}\right)-2366.51 \delta _i\right]+\nonumber\\
   && 3253.9\left(0.888 \delta _i+\frac{\delta
   '_i}{2}\right)\bigg\} w^{3/2}+0.00925022\times\nonumber\\
   &&\left\{5094.72 \delta _i -1.66667\right.\times \nonumber\\
    &&\left.\left[2971.92 \delta _i
   -1877.38 \left(0.888 \delta _i +\frac{\delta
   '_i}{2}\right)\right]\right.+\nonumber\\
   &&2.22222 \left[198.128 \delta _i
   -125.159 \left(0.888 \delta _i+\frac{\delta
   '_i}{2}\right)\right]-\nonumber\\
   &&\left.3218.36 \left(0.888 \delta _i
   +\frac{\delta '_i}{2}\right)\right\}
    w^{5/3}+
   O\left(w^{7/3}\right), w\rightarrow0.\nonumber\\
   \end{eqnarray}
Near the initial state $w=1/2$, the density contrast can be approximated as
\begin{eqnarray}
\delta \left( w\right)  &\approx &\delta _{i}+
\delta _{i}^{\prime }\left( w-\frac{1}{2}\right) +2.39916\times 10^{-6}\times \nonumber\\
&&\left( 756436\delta
_{i}-578905\delta _{i}^{\prime }\right) \left( w-\frac{1}{2}\right) ^{2}+ \nonumber\\
&&2.97184\left( -1.10272\delta _{i}+0.864771\delta _{i}^{\prime }\right)
\left( w-\frac{1}{2}\right) ^{3}+\nonumber\\
&&O\left( w-\frac{1}{2}\right) ^{4}, w\rightarrow 1/2.
\end{eqnarray}

\begin{figure}
\centering
\includegraphics[width=8.15cm]{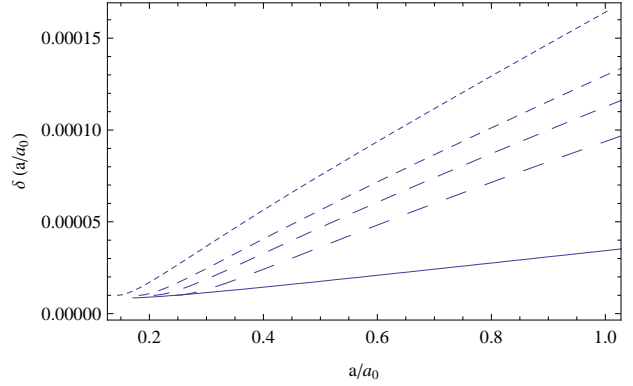}
\caption{Evolution of the density perturbations $\delta \left(a/a_0\right)$ as a function of $a/a_0$ of the pressureless dark matter, described by the Einstein-de Sitter metric (solid curve), and of the Bose-Einstein condensate dark matter, for different values of $\rho _0$: $\rho _0=10^{-3}$ (dotted curve), $\rho _0=2\times 10^{-3}$ (dashed curve), $\rho _0=3\times 10^{-3}$ (long-dashed curve), and $\rho _0=5\times 10^{-3}$ (ultra-long dashed curve). In all cases the initial conditions are $\delta \left(3\rho _0\right)^{1/3}=10^{-5}$ and $\delta ^{\prime} \left(3\rho _0\right)^{1/3}=10^{-5}$, respectively. }
\label{fig2}
\end{figure}

\begin{figure}
\centering
\includegraphics[width=8.15cm]{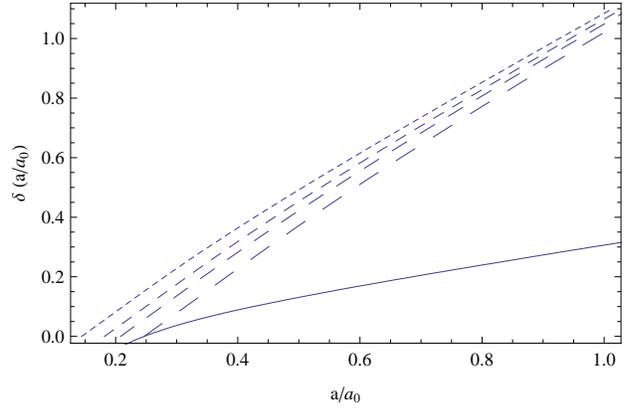}
\caption{Evolution of the density perturbations $\delta \left(a/a_0\right)$ as a function of $a/a_0$ of the pressureless dark matter, described by the Einstein-de Sitter metric (solid curve), and of the Bose-Einstein condensate dark matter, for different values of $\rho _0$: $\rho _0=10^{-3}$ (dotted curve), $\rho _0=2\times 10^{-3}$ (dashed curve), $\rho _0=3\times 10^{-3}$ (long-dashed curve), and $\rho _0=5\times 10^{-3}$ (ultra-long dashed curve). In all cases the initial conditions are $\delta \left(3\rho _0\right)^{1/3}=10^{-5}$ and $\delta ^{\prime} \left(3\rho _0\right)^{1/3}=1.5$, respectively. }
\label{fig3}
\end{figure}

Since Eq.~(\ref{pertnum}) is valid only in the linear regime of small perturbations, we assume that the initial value of the perturbation, $\delta \left(a_i/a_0\right)$, occurring for a value $a=a_i$ of the scale factor, satisfies the condition $\delta \left(a_i/a_0\right)<<1$. Since the equation describing the perturbations is a second order differential equation, two initial values have to be given, one for the initial perturbation $\delta \left(a_i/a_0\right)$, and one for the initial rate of evolution of the perturbation, $\delta ^{\prime}\left(a_i/a_0\right)$. We consider two cases, namely,  the case of a perturbation with an initial low evolution rate, of the order of $\delta ^{\prime} \left(a_i/a_0\right)=10^{-5}$, and the case of a perturbation with a very high initial evolution rate, $\delta ^{\prime} \left(a_i/a_0\right)=1.5$, respectively. The comparison between the evolution of the evolution of the linear perturbations for pressureless dark matter in an expanding Einstein - de Sitter cosmological background, and the evolution of the density perturbations in a Bose-Einstein condensate dark matter dominated Universe is presented in Figs.~\ref{fig2} and \ref{fig3}, respectively.

As one can see from Figs.~\ref{fig2} and \ref{fig3}, for all initial conditions, in the case of the  Bose-Einstein, for a given $a$,  the amplitude of the density contrast is higher as compared to the case of the standard dark matter model. The condensate enters more rapidly in the non-linear phase ($\delta >>1$) than the pressureless dark matter. Thus the presence of the Bose-Einstein condensate dark matter can significantly accelerate the process of cosmic structure formation.

\section{Discussions and final remarks}\label{sect7}

In the present paper we have considered the global cosmological evolution and the evolution of the small cosmological perturbation in a Bose-Einstein dark matter condensate. The basic equation describing the evolution of the small perturbations in the Post-Newtonian regime was obtained, and its solutions have been studied by using both analytical and numerical methods. The evolution of the density perturbations of the condensate has been compared to the evolution of the small cosmological perturbations in a pressureless fluid evolving in an Einstein-de Sitter cosmological background. Depending on the numerical values of the physical parameters describing the condensate (the mass of the particle and the scattering length, respectively), significant differences could appear in the evolution of the Bose-Einstein condensate dark matter halos, as compared to the standard pressureless dark matter models. These differences appear at both the level of the global cosmological evolution, and of the behavior of the small perturbations in the dark matter fluid, and they could have fundamental implications for the formation of the large-scale structure in the Universe.

 One of the most important problems present day cosmology faces is the problem of the galaxy formation. To explain galaxy formation the evolution of the linear density and temperature perturbations in a Universe with dark matter, baryons, and radiation must be computed. For pressureless dark matter the evolution of the perturbations of all cosmic components,  from cosmic recombination until the epoch of the first galaxies, was obtained in \citet{Na}. The evolution of sub-horizon linear perturbations can be  described by two coupled, second-order differential
equations, with the pressureless dark matter interacting gravitationally with itself, and with the baryons, while the baryons experience both gravity and pressure. Starting from very low values on sub-horizon
scales, the baryon density perturbations gradually approach those in the dark matter, and  the temperature perturbations approach the value expected for an adiabatic gas. The presence of the baryons does not modify significantly the evolution of the dark matter perturbations. By including the effect of the BEC pressure in the perturbations equations for baryons and dark matter, a more general (and realistic) description of the galaxy formation process can be obtained. The presence of the BEC modifies the dynamical evolution  of the baryons, and the growth of linear perturbations, which provide the initial conditions for the formation of galaxies. In the BEC condensate model the dark matter perturbations grow more rapidly than in the standard cosmology, and therefore this could lead to a much faster growth rate of the baryonic perturbations, accelerating the galaxy formation process.

A major recent experimental advance in the study of the Bose-Einstein condensation processes was the observation of the collapse and subsequent explosion of the condensates \citep{ryb04}.  A dynamical study of an attractive
$^{85}$Rb BEC in an axially symmetric trap was done, where the interatomic interaction was manipulated by changing the external magnetic field, thus exploiting a nearby Feshbach resonance. In the vicinity of a Feshbach resonance the atomic scattering length a can be varied over a huge range, by adjusting an external magnetic field. Consequently, the sign of the scattering length is changed, thus transforming a repulsive condensate of $^{85}$Rb atoms into an attractive one, which naturally evolves into a collapsing and exploding condensate. From a simple physical point of view the collapse of the Bose-Einstein condensates can be described as follows. When the number of particles becomes sufficiently large, so that $N>N_c$, where $N_c$ is a critical number, the attractive inter-particle energy overcomes the quantum pressure, and the condensate implodes. In the course of the implosion stage, the density of particles increases in the small vicinity of the trap center. When it approaches a certain critical value, a fraction of the particles gets expelled. In a time period of an order of few milliseconds, the condensate again
stabilizes. There are two observable components at the final stage of the collapse: remnant and burst particles. The remnant particles are those which remain in the condensate. The burst particles have an energy much larger than that of the condensed particles. There is also a fraction of particles, which is not observable. This fraction is usually referred to as the missing particles.

The scattering length $l_a$ is defined as the zero-energy limit of
the scattering amplitude $f$~\citep{Da99}. Depending on the spin
dependence of the underlying particle interaction, the scattering
length may in general be also spin dependent. The spin independent
part of the quantity is referred to as the coherent scattering
length $l_a$. The scattering lengths can be obtained for some
systems in the laboratory, but for dark matter it is unknown.
Another essential parameter is the mass $m$ of the condensate
particle, which, due to the lack of information about the physical
nature of the dark matter, is a free parameter, which must be
constrained by observations. Due to the lack of any physical information about the numerical values of these two fundamental parameters, in the numerical estimations performed in the present paper we have given different numerical values to a combination of these two basic quantities.

Since Bose-Einstein condensates are less stable with respect to perturbations than usual non-condensate matter, we expect that in such a condensate evolution of the perturbations and the subsequent collapse could take place much faster than in the usual non-condensed matter. This would strongly affect the formation of the large scale structure in the early Universe.  In this paper we have provided some
basic theoretical tools necessary for the in depth comparison of the predictions of the condensate model and of the observational results.

\section*{Acknowledgments}

I would like to thank to the anonymous referee for comments and suggestions that helped me to considerably improve the manuscript. The work described in this paper was fully supported by a grant from the Research Grants Council of the Hong Kong Special Administrative Region, China.

\end{document}